# Weighted delay-and-sum beamforming guided by visual tracking for human-robot interaction

*José Novoa[1], Rodrigo Mahu[1], Alejandro Díaz[1], Jorge Wuth[1], Richard Stern[2], Nestor Becerra Yoma[1]*

[1]Speech Process. and Transm. Lab., Elec. Eng. Dept., U. de Chile, Santiago, Chile.
[2] ECE Dept. and Language Technologies Institute, CMU, Pittsburgh, PA 15213, USA.

`nbecerra@ing.uchile.cl`

**Abstract**

This paper describes the integration of weighted delay-and-sum beamforming with speech source localization using image processing and robot head visual servoing for source tracking. We take into consideration the fact that the directivity gain provided by the beamforming depends on the angular distance between its main lobe and the main response axis of the microphone array. A visual servoing scheme is used to reduce the angular distance between the center of the video frame of a robot camera and a target object. Additionally, the beamforming strategy presented combines two information sources: the direction of the target object obtained with image processing and the audio signals provided by a microphone array. These sources of information were integrated by making use of a weighted delay-and-sum beamforming method. Experiments were carried out with a real mobile robotic testbed built with a PR2 robot. Static and dynamic robot head as well as the use of one and two external noise sources were considered. The results presented here show that the appropriate integration of visual source tracking with visual servoing and a beamforming method can lead to a reduction in WER as high as 34% compared to beamforming alone.

## 1. Introduction

### 1.1. Automatic speech recognition and human-robot interaction

Human beings communicate through speech without much effort, even in the most unfavorable conditions. Our speech recognition and noise filtering capabilities perform better than any system implemented for this purpose [1]. Human-robot interaction (HRI) commonly involves automatic speech recognition (ASR). Nevertheless, most realistic HRI scenarios are not characterized by a noiseless acoustic channel. Even the noise produced by the motors and mechanisms of the robot affects the performance of the ASR systems, especially when the servos move near the microphone. Mechanical movement produces non-stationary noise that depends on the interaction. In [2] the present authors proposed the integration of ASR for HRI applications considering the acoustic environment with noise produced by robot motors. In this paper we replace the classic black box integration of ASR systems with a multimodal strategy.

### 1.2. Weighted delay-and-sum and microphone arrays

A microphone array is an arbitrary number of microphones working together. The use of a microphone array to perform beamforming can reduce the effect of reverberation and noise by suppressing the non-direct path acoustic signals [3]. For example, the Microsoft Kinect, which is widely used in HRI applications, has a 4-channel linear microphone array (see Fig. 1), along with standard RGB and depth cameras.

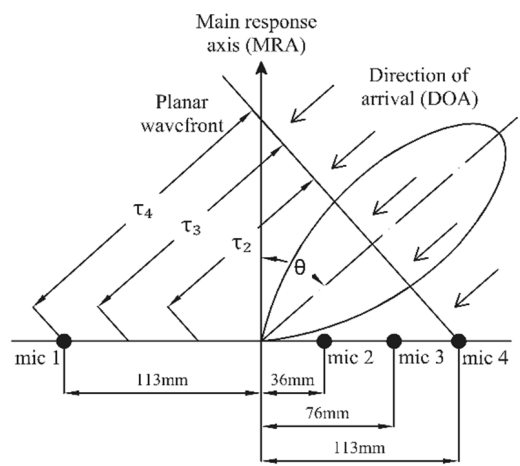

Figure 1: *The geometry of the Microsoft Kinect microphone array, the channel delays, the main response axis (MRA), the direction of arrival (DOA) and a hypothetical beamformed lobe are illustrated in this figure.*

Delay-and-sum is a well-known technique for this purpose, given a known direction of arrival (DOA) and time delays. It sums the delayed signal depending on the direction of arrival of the sound waves. This produces destructive interference in all directions but the direction of arrival. There are different microphone arrays shapes and types, so each singular microphone may capture the same sound differently. Weighted-delay-and-sum is a generalized form of delay-and-sum, with signal samples $x_n(k)$ from each microphone $M_n$ delayed by $\tau_n$ samples, multiplied by weights $w_n(k)$ and then summed. By doing this, the output signal $y(k)$ in the discrete-time domain corresponds to:

$$y(k) = \sum_{n=0}^{N-1} w_n \cdot x_n(k - \tau_n) \quad (1)$$

A planar wavefront can be assumed if the distance between the microphone array and the sound source is larger than 5-10 times the length of the array [4]. Consequently, the delay for each microphone is given by

$$\tau_n = \frac{\Delta_n \cdot \sin\theta}{c} \quad (2)$$

where $\Delta_n$ is the distance between the microphone $n$ and the reference microphone. The angle of incidence (AOI) is $\theta$, which corresponds to the angle between the main response axis (MRA) and DOA, finally c is the propagation speed of sound in the medium [5].

BeamformIt is a tool described in [6] that performs weighted-delay-and-sum beamforming for an arbitrary microphone-array designed to operate in the context of meetings or conferences. It consists of four stages: the first stage applies a Wiener filter to each channel, which seeks to increase the SNR in that channel. The next stage extracts information from the inputs, and determines some parameters used in later stages. This second stage consists of 4 sub-stages, the first being the determination of the channel that best represents the acoustics of the room, which is called the 'reference channel.' This is accomplished by calculating the windowed mutual correlations among the signals from the microphones. The next sub-step consists of determining an overall weight associated with the channels to take better advantage of the dynamic range of the signal resulting from beamforming. The fourth sub-step considers N candidates (typically 4) of TDOA's (time direction of arrival) for each channel. This is implemented by maximizing the GCC-PHAT (Generalized Cross-correlation with Phase Transform) between each signal and the previously-determined reference channel. The greatest N values are chosen and saved for later stages. After obtaining the N candidates for measurement from each microphone, a selection is made by comparing with noise thresholds in the first instance, and then using Viterbi to select the optimum combination of delays to be used in each channel. This consists of the third stage of the algorithm. The fourth and last stage corresponds to the generation of the output signal. The first thing that is done is to estimate the weights of each channel, which compensate for the possible hardware differences of each microphone, and the response to the impulse observed by each one of them. The weights are initially estimated as equal, and are adapted over time, based on the average cross-correlation between each channel and all others, after applying the corresponding delays. This average cross-correlation value is also used to eliminate microphones with poor quality signals, if they are below a previously-defined threshold. Finally, the signals of each channel are added to obtain the output signal.

### 1.3. Beamforming directivity gain

It is known that the reliability of beamforming decreases with unfavorable environmental conditions causing problems such as ghost sources, while multiple sources appear to be in other directions, due to successive reflections in walls [7]. Moreover, the sensitivity of a microphone array decreases when the AOI increases [4], [8]. In practice, sensitivity decreases dramatically [9] with adaptive beamforming, when the target moves away from the MRA.

### 1.4. Image tracking and visual servoing

Object tracking means following the movement of a detected object while it moves around frames in a video. There are multiple methods that can be used for object tracking, one of them is "You Only Look Once" (YOLO), which is a real-time object detection system, to detect and track objects. It uses deep learning and convolutional neural networks. And is widely used for multiple purposes, such as the detection and tracking of marine organisms [10], and by the winning participants of Robocup 2018 in Montreal [11]. The information extracted from object detection, can also be used for visual servoing, which refers to controlling a servo motor using visual information in a feedback loop [12].

### 1.5. About this paper

This paper presents an ASR system for HRI that uses the information resulting from image processing to control the real-time orientation of a linear microphone array MRA to achieve better performance in realistic environments, where the effect of noise and reverberation makes distant speech recognition a challenging task. For doing this, a visual servoing scheme is used to reduce AOI. Two information sources were integrated in the multimodal beamforming strategy that is described: the direction of the target object obtained with image processing and the audio signals provided by a microphone array. These sources of information were combined using the weighted delay-and-sum performed by BeamformIt and are compared to classical delay-and-sum beamforming.

## 2. HRI testbed

The proposed testbed is a time-varying and noisy acoustic channel similar to the HRI scenario proposed in [2] which is representative of many situations where humans and robot might interact collaboratively.

### 2.1. Robot movements

Fig. 2 shows the HRI scenario used in this paper. Where the PR2 robot performs periodic lateral displacements from position P1 to position P3 reaching a maximum velocity of 0.45m/s, applying an acceleration and deceleration when reaching terminal positions. Meanwhile, the robot head may perform rotational movements following the scheme described in Section 2.4

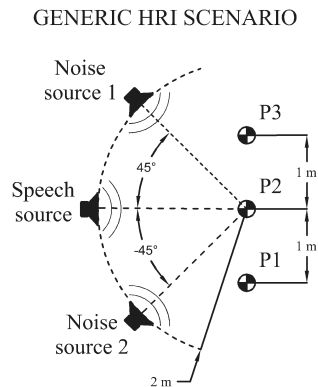

Figure 2: *Scenario layout used in the proposed testbed. The PR2 robot performs translational movement between the positions P1 and P3 while the head is kept in two conditions: a) static; and b) dynamic, attempting to reduce the angle of incidence (AOI) employing visual servoing.*

### 2.2. Target speech source and external noise source

The speech source was a TANNOY 501a loudspeaker, located two meters away from position P2 as shown in Fig. 2. It reproduced 330 clean testing utterances from the Aurora-4 database [13]. To avoid interference between utterances, the playback was paused for five seconds between each utterance.

Along with the speech source, one or two external noise sources were used to make the conditions more challenging. The noise reproduced by the loudspeaker was non-stationary restaurant noise at an SNR of 5 dB. Also, noise produced by the robot motors, fans and wheels was always present.

### 2.3. Object detection and image tracking

For object detection and video tracking, we ran YOLO [14] on a GeForce GTX 1080 GPU with a previously-trained CNN. The system runs on Darknet, an open source neural network framework written in C and CUDA. YOLO was employed to recognize the target speech source and to estimate its angular position within the frame with respect to the image center. This information can be used to separate a target speech source from other noise sources using spatial filtering.

### 2.4. Beamforming and visual servoing

In this paper, the use of weighted delay-and-sum beamforming with the Kinect microphone array was explored and evaluated with the mobile robotic testbed described above. The visual-based beamforming strategy proposed in this paper to implement the weighted delay-and-sum beamforming, which makes use of constructive interference to direct the main lobe to the desired direction considering knowledge of the channel phase shifts [3] (see (1) and (2)). Equation (2) uses AOI to compute the time delays for each channel. We estimate this angle from the image processing performed by YOLO, considering the speech target source angular position as the direction of arrival and assuming a planar wave front. We used the Kinect RGB 0.3MP camera mounted on the PR2 robot head to control an end-effector of a robot with data extracted from visual sensors in a feedback loop, which is called visual servoing (*e.g.* [12]). The flowchart of the visual servoing algorithm used in this research is presented in Fig. 3. Starting from the "PR2 Kinect camera" block in the diagram, the robot camera streams video in real time to the GPU server for image processing by YOLO. The GPU server sends the target coordinates describing where to move the head back to the PR2 robot where the movement is planned. If the target is not centered, an action is performed on the robot head. After that, the system waits for a new transmitted frame and the cycle is repeated. By doing so, the robot head is actively oriented towards the speech source. In this paper, the speech source detection was carried out by detecting a personalized template inspired by the stop sign, which is one of the many object classes that YOLO had been trained to detect (see Fig. 4). The template was placed on the top of the studio loudspeaker employed to reproduce speech.

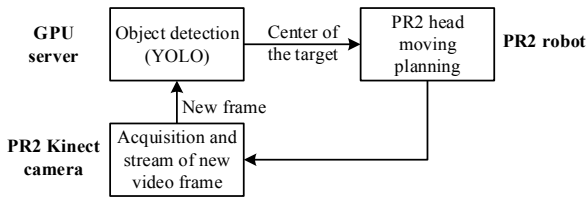

Figure 3: *Flowchart of the visual servoing scheme presented here.*

## 3. Testing databases

In this section four testing databases recorded with the Microsoft Kinect available on top of the PR2 robot in the interaction scenario shown in Fig. 2 are presented. The Kinect SDK library available on the Microsoft website [15] was employed for the audio recording, where the four channels of the Kinect was saved separately as well as the beamformed signal. The datasets were recorded with the robot performing periodic lateral displacements between positions P1 and P3 as described in Section 2. The AOI of the target speech source is obtained, as well as the current angular position of the head "published" in ROS (Robot Operating System) were saved. Two conditions were considered for robot head movement in the recordings: keeping the robot head fixed at 0°, *i.e.* perpendicular to the translational movement made by the base of the robot; or a robot head gaze that follows the speech source based on image tracking and the proposed visual servoing scheme as described in Section 2. For each head movement condition, two external noise sources conditions were considered: employing only Noise source 1 as described in Section 2, or employing Noise source 1 and Noise source 2. Table 1 summarizes the different recording conditions:

Table 1: *Datasets recording conditions.*

| Dataset | Head condition | Noise sources |
| --- | --- | --- |
| NST-1 | Fixed at 0° | Noise source 1 |
| NST-2 | Fixed at 0° | Noise source 1 and 2 |
| VbST-1 | Following speech source | Noise source 1 |
| VbST-2 | Following speech source | Noise source 1 and 2 |

## 4. Automatic speech recognition

Speech recognition experiments were performed with a DNN-HMM ASR using the Kaldi Speech Recognition Toolkit [16]. To build a DNN-HMM system with Kaldi, first a GMM-HMM is trained according to the tri2b Kaldi Aurora 4 recipe with the training data described below. The GMM-HMM system was trained using MFCC features, linear discriminant analysis (LDA), and maximum likelihood linear transforms (MLLT). A monophone system was trained first; then, the alignments from that system were employed to generate an initial triphone system; finally, the triphone alignments were employed to train the final triphone system. Then, the GMM in the trained GMM-HMM system was replaced with a DNN composed of seven hidden layers and 2048 units per layer each, and the input considers a context window of 11 frames. The number of units of the output DNN layer is equal to the number of Gaussians in the corresponding GMM-HMM system. For decoding, the standard 5K lexicon and trigram language model from the DARPA Wall Street Journal database (WSJ) were used [17].

### 4.1. Training data

The training dataset was generated considering 33 four-channel impulse responses (IRs) in the same way as the Environment-based Training (EbT) data in [2]. This IRs were estimated with the Kinect MRA oriented to 11 different angles between 150° and -150° with the microphone array located at one, two, and three meters from the speech source. Similarly to EbT, 25% of the clean data from the Aurora-4 database were convolved with the IR estimated with the MRA of the microphone array pointing to the speech source at one meter from the speech source. Noise consisting of robot noise plus uncorrelated external restaurant noise was added to the remaining 75% of the clean data. In contrast to the EbT data (which were combined at a random SNR between -5 dB and 5 dB), the SNRs of our data were between 10 and 20 dB, as measured using the FaNT tool [18].

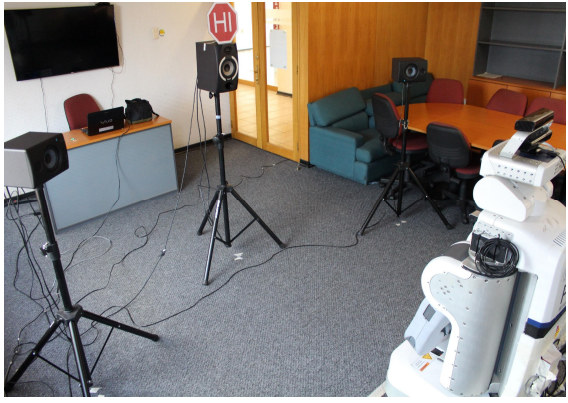

Figure 4: *HRI testbed with the PR2 robot showing the target speech source and the external noise sources.*

## 5. Results and discussion

The average absolute AOI obtained for the robot head fixed at 0° while performing translational movement is 15.7°. In contrast, when the robot head gaze followed the speech source the average absolute AOI is equal to 4.5°, which corresponds to a reduction of 71%.

Figure 5 summarizes the results obtained using three beamforming strategies: BeamformIt, which corresponds to the weighted sum and delay scheme presented in [6]; B+AOI, which corresponds to BeamformIt with the delays computed from the AOI estimated by visual tracking; and, D&S+AOI, that corresponds to the classical delay and sum with the delays computed from the AOI estimated by visual tracking. As seen in Fig. 5, the WERs obtained with B+AOI were 6.0% and 9.4% lower than the one with BeamformIt with NST-1 and VbST-1, respectively. These reductions are due to the fact that in B+AOI the angle of incidence is estimated by visual tracking, which is sensitive to neither reverberation nor external noise sources. Additionally, the greater reduction in VbST-1 occurs because the visual servoing scheme decreases the average absolute AOI and improves the sensitivity of the generated beamforming as explained above. This result suggests that both image source tracking and visual servoing enhance the beamforming process in a complementary fashion. When considering two external noise sources, the WERs obtained with B+AOI were 0.9% and 5.2% lower than results obtained using BeamformIt in the NST-2 and VbST-2 conditions, respectively.

Additionally, the highly dynamic scenario considered in this paper requires a reduction in the duration of the analysis window employed by BeamformIt. Unfortunately, a narrower analysis window leads to worse correlation estimation, and hence the reference channel and channel weight estimates are degraded. On the other hand, given the closeness and similarity of the Kinect microphones, it seems sensible to weight all the channels uniformly and pick any of them as a reference. The resulting beamforming corresponds to the classical delay-and-sum scheme, which in turns degrades recognition accuracy substantially compared to the original weighted delay-and-sum. However, Fig. 5 describes delay-and-sum beamforming in combination with the AOI estimated with visual tracking, D&S+AOI. The analysis window was reduced from the original 0.5s to 0.05s. According to Fig. 5, the WERs obtained with D&S+AOI were 12.2%, 15.7%, 8.0%, 9.7% lower than those obtained using BeamformIt with the NST-1, VbST-1, NST-2 and VbST-2 data, respectively.

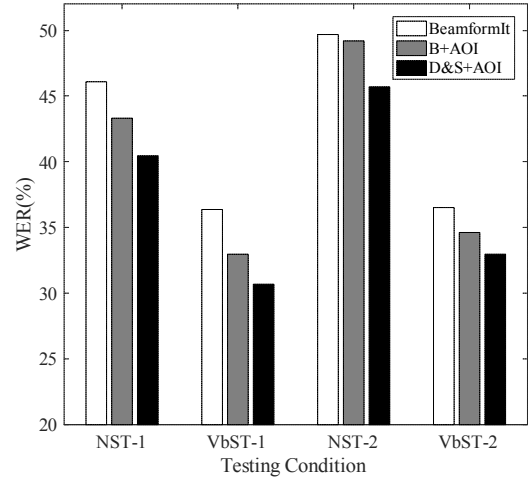

Figure 5: *Schematic diagram of speech production.*

## 6. Conclusions

This paper describes the integration of the target speech source direction estimation using image processing for multimodal beamforming and a visual servoing scheme for source tracking was proposed to improve the ASR accuracy in HRI indoor environments. This strategy combines information from audio signals and image processing, with robot head mobility. The proposed scheme was evaluated in realistic and challenging HRI testing conditions, and compared with the traditional weighted delay-and-sum beamforming methods. For this purpose, we built a real mobile robotic testbed with a PR2 robot and external noise sources. The results obtained here show that the use of multiple information sources with the robot head mobility leads to dramatic improvements in the ASR accuracy. Also, the channel weighting scheme seems redundant in the scenario considered here that needed to reduce the beamforming analysis window. As expected, visual source tracking combined with delay-and-sum beamforming can lead to a reduction in WER equal to 12% when compared with the original weighted delay-and-sum beamforming. Visual servoing combined with visual tracking and delay-and-sum beamforming can lead to an average reduction in WER as high as 33.5% when compared with the original weighted delay-and-sum beamforming scheme. Moreover, image source tracking and visual servoing enhance the beamforming process in a complementary fashion. Improving visual servoing with motion-estimation methodologies, improving speaker tracking with image-detection techniques, integration with more sophisticated beamforming schemes and evaluation with higher computation capability GPUs can be considered in future research. Finally, our strategy is applicable to any HRI environment where audio and visual information are both available.

## 7. Acknowledgements

The research reported here was funded by Grants Conicyt-Fondecyt 1151306 and ONRG N°62909-17-1-2002.